# Adjustment of electric field intensity by dielectric metamolecule hybridization model

Haohua Li,[a] Xiaobo Wang, [a] Ji Zhou [a*]

a. State Key Laboratory of New Ceramics and Fine Processing, School of Materials Science and Engineering, Tsinghua University, Beijing 100084, China.
*Correspondence: zhouji@tsinghua.edu.cn

**Abstract**:
In this paper, we report on achieving hybridization effect in a Mie-based dielectric metamolecule and provide their physically intuitive picture. The hybridization results in the splitting of the initial overlapping resonance dips, thus leading to two new collective resonance modes. It can be observed via the displacement electric field distribution that the two modes behave as the in-phase and out-of-phase oscillation of two meta-atoms, thus enhancing and diminishing the intensity of electric field at the gap between two meta-atoms. Moreover, since the two hybridized modes are caused by the interaction effect, the enhancement amplitude can be adjusted by several coupling factors, like gap distances and spot positions. Taking advantage of the two collective modes of the dielectric metamolecule, certain locations in the metamolecule can be applied as heating zones in microwave band and signal amplifier or shielding zones in communication fields.

**1. Introduction**

Veselago put forward the concept of metamaterial from Maxwell equation in 1967[1], theoretically predicted a series of novel electromagnetic behaviors[2-5]. Since the size of the single units and separation between unit cells within a metamaterial are smaller than the working wavelength, the interaction between the meta-atoms initiates the coupling hybridization phenomenon and corresponding applications [6-10]. Inspired by the molecular orbital diagram and energy theory, scientists explain the energy level splitting of a coupled system by applying various hybridization models [11-14]. Halas and Nordlander first introduced the hybridization model to the nanoshell structure[7], whose resonance frequency can be attributed to the interaction of a nanosphere and a nanocavity. Verellen, N proposed a gold nanocross cavity structure assembled using a nanocross and a nanorod[15]. Owing to interference of the quadrupolar and dipolar modes of X and I elements, the whole resonance of the XI structure hybridizes into a bonding quadrupolar X-mode and a subradiant bonding dipolar mode. Hybridized modes in metamolecule always lead to enhanced electric fields, thus bringing out various applications, such as improved fluorescence emission[16], surface-enhanced Raman scattering[17,18]. Pablo Alonso-González put forward an infrared dimer nanoantenna that produces enhanced and suppressed electric fields in hybridized modes[19]. Rostam Moradian proposed a bimetallic core-shell structure to realize surface enhanced Raman scattering by means of hybridized resonance modes[20].

However, most coupling hybridization models focus on optical and terahertz wavelength[9,10,12,21,22]; compared with those, microwave hybridization have special uses in communication fields. Besides, compared to metal, dielectric metamaterials have advantages of low losses, simple structures and active regulatory[23-26]. In this paper, we discuss a hybridization model in microwave band composed of dielectric meta-atoms. By means of hybridization between two dielectric meta-atoms, interference occurs at their initially common Mie-resonance frequency, which leads to a transparent window and two new collective resonance modes being produced. Since dielectric materials produce the first-order Mie-resonance by excitation of microwave, behaving as magnetic loops formed of orientation arrangement of displacement electric currents. In the two collective modes, the displacement electric current oscillated as "in-phase" mode and "out-of-phase" mode in two meta-atoms, respectively, thus strengthening and weakening the electric fields between them. By utilizing the enhancement and diminishment interaction, we can obtain electric fields with various intensities by adjusting several factors that affect coupling effect, like gap distances, spot positions and permittivities of meta-atoms. Here we reveal how displacement electric field distributes in the meta-atoms in two hybridized collective modes, phenomenally explaining the coupling manner of "heteronuclear diatomic metamolecule". Besides, we put forward how to obtain strengthened or weakened electric fields with adjustable intensities by utilizing these two collective modes, which is promising for amplifying or shielding signals in communication fields.

## 2. Design, simulation and experiment

As figure 1 shows, the dielectric metamolecule is composed of $CaTiO_3$ (CTO) and $SrTiO_3$ (STO) meta-atoms, shaping into cuboids. The complex permittivities of $CaTiO_3$ and $SrTiO_3$ are $\varepsilon_r$=160, $\tan\delta$ =0.001 and $\varepsilon_r$=317, $\tan\delta$ =0.003, respectively. And the dimensions of the CTO and STO meta-atoms are carefully designed to be 2 × 1.8 × 2 $mm^3$ and 1 × 2 × 2 $mm^3$, respectively, matching with their permittivity so that they can have nearly the same first-order Mie resonance frequency according to equation (1):

$$f = \frac{\theta c}{2\pi r\sqrt{\varepsilon\mu}} \qquad (1)$$

where $\theta$ is a coefficient that approximates to $\pi$, c is light velocity in vacuum, $\mu$ equals to 1 for non-magnetic materials, and r and $\varepsilon$ represent for dimension and permittivity of the dielectric meta-atom.

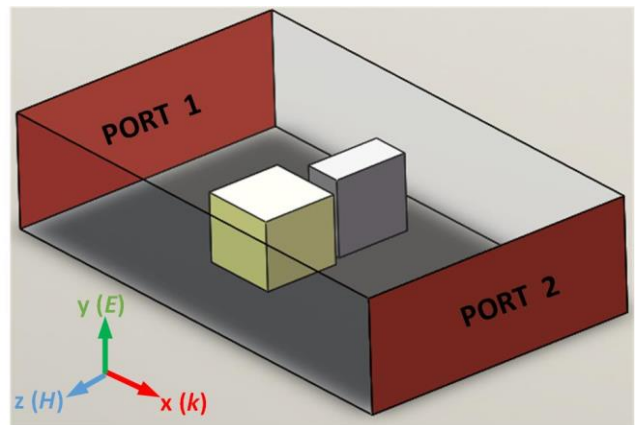

Figure 1. Schematic diagram of metamolecule in a waveguide. The incident wave propagates along the x-axis, the magnetic field and the electric field is along the z-axis and y-axis, respectively.

Simulation is performed using the commercial software package CST to calculate the transmission spectra of both the single meta-atoms and the assembled metamolecule. As dash curves in figure 2(a) shows, the two meta-atoms produce dramatic transmission dips at very similar frequencies around 10.48GHz for an incident electromagnetic wave, and as such the two dips nearly overlap. This means that the meta-atoms with their specifically designed dimensions and permittivities are opaque to the incident electromagnetic wave at a coincidental frequency. The displacement electric current at the resonance frequency oscillate along the entirety of dielectric cuboids, distributing itself so as to form a magnetic dipole, indicating that the first-order Mie resonance occur at the nearly overlapping frequency found for the meta-atoms. The metamolecule described in Figure 1 shows that two meta-atoms are arranged side-by-side along the z-axis, parallel to the direction of the magnetic field of the incident wave. The distance between the two meta-atoms is set to be 0.6mm, which is much smaller than the working wavelength, thus the coupling between the two meta-atoms is strong enough to induce redistribution of displacement electric fields in metamolecule. Solid curves in figure 2 (a) show the transmission response $S_{21}$ of the metamolecule when an electromagnetic wave is incident on it at 9.5-11GHz. We can see that a typical hybridization induced transparency (HIT) phenomenon appears. An electromagnetic transparent window replaces the original intrinsic resonance dips of the meta-atoms; in addition, two new transmission dips of the metamolecule emerge at 10.16GHz and 10.727GHz. It is worth mentioning that the amplitudes of the two dips are −15.02dB and −1.16dB; while the initially most opaque frequency becomes transparent at amplitude of −0.24dB.

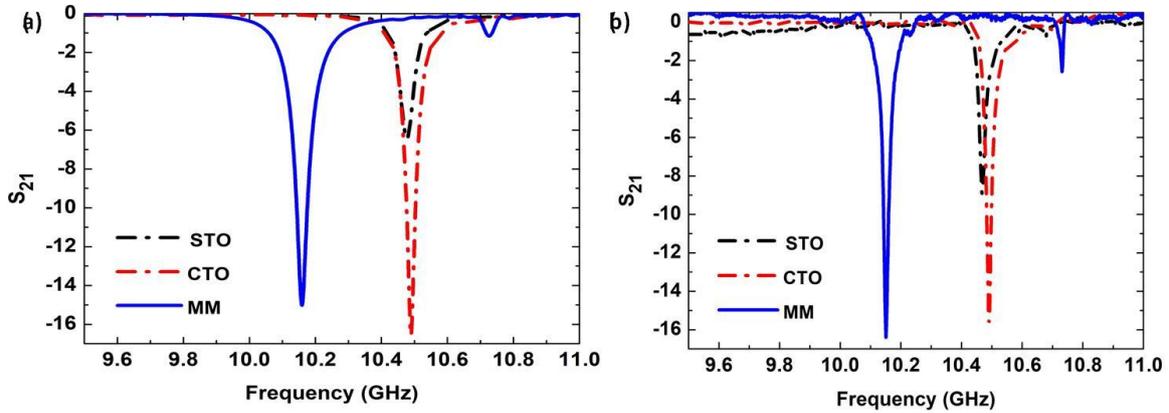

Figure 2. Simulated (a) and measured (b) $S_{21}$ curves of CTO and STO meta-atoms (dashed lines) and the metamolecule (solid lines).

The $S_{21}$ curves obtained in the experiment are measured by a microwave system composed of an EIA Agilent N5230C vector network analyzer and XB-SW90 rectangular waveguides. Dash curves in figure 2(b) show the transmission spectra of the meta-atoms. The transmission dips of the CTO and STO cuboids appear at 10.49GHz and 10.47GHz, respectively; as such, they agree well with the results of the simulation. Solid curves in figure 2(b) depict the transmission spectra of the metamolecule. Two dips can be seen at 10.15GHz and 10.73GHz; these findings agree well with the results obtained from the simulation.

## 3. Results and discussion

Since metamolecule is composed of dielectric meta-atoms, their first-order Mie resonance is caused by displacement electric current as a magnetic loop. To fully explore the mechanism behind the phenomenon of hybridization induced transparency (HIT), the distribution of the displacement electric field of the metamolecule is observed at the frequencies of both the collective resonance dips and the transparent window. As figure 3(a) shows, at the red-shifted collective resonance dip of 10.16GHz in the simulation, two magnetic dipoles appear in both meta-atoms; they oscillate in the parallel direction and are found to be in an in-phase mode. The other collective mode of the metamolecule is blue-shifted to 10.727GHz and is visible as a slight dip. As figure 3(b) shows, the displacement electric fields in the two meta-atoms oscillate in anti-parallel directions and are therefore considered to be in an out-of-phase mode. Because it is weakened by the counteracting effect of the anti-parallel oscillation, the amplitude of the resonance at 10.727GHz is much weaker than the former one. It should be noted that both of the induced magnetic fields in the meta-atoms are parallel to the magnetic field of the incident electromagnetic wave, which indicates that the new collective modes occur in response to the application of the external magnetic field. As for the initial resonance frequency of the meta-atoms at ca. 10.48GHz, it can be seen from figure 3(c) that the electric field spreads through the waveguide freely without localized in dielectrics; as a result, the electromagnetic transparency is induced.

Like "heteronuclear diatomic molecules" and their constituents, this can be viewed as a hybridization model for the dielectric metamolecule, as described in figure 3(d). Meta-atoms initially share the same Mie-resonance frequency due to their intrinsic properties. Owing to the strong coupling effect, interaction occurs at their overlapping orbits. The metamolecule generates new collective orbits, which displays as two new resonance dips. Coupled meta-atoms combine as amended "dielectric diatomic molecules"; the red-shifted and blue-shifted collective modes can be regarded as a low-energy "bonding" mode and a high-energy "anti-bonding" mode. Thus, the metamolecule with two collective modes produce an asymmetric transmission spectrum.

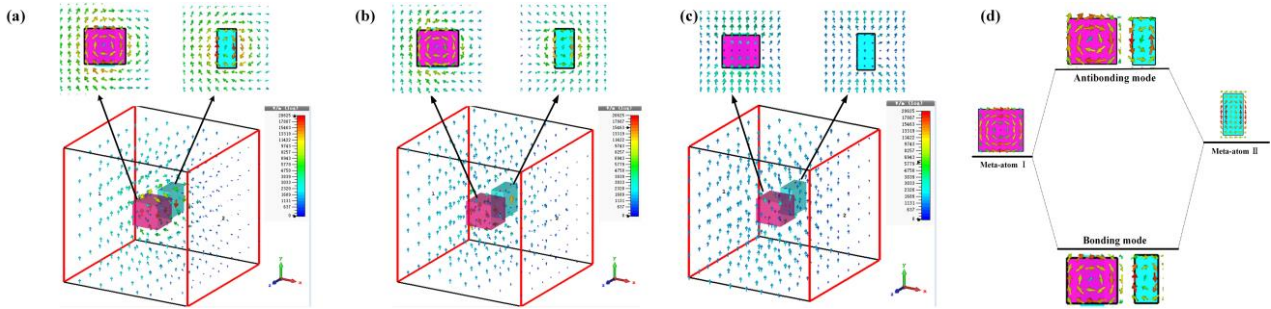

Figure 3. Electric field distribution in the metamolecule at (a) 10.16GHz, (b) 10.727GHz and (c)10.48 GHz; the illustrations show the electric field of each meta-atom perpendicular to z-axis in the central section of z=1.3mm and z=-1.3mm, respectively. (d) Analog hybridization picture of the metamolecule with an electric field distribution at each resonance frequency.

From the dielectric metamolecule hybridization model, it can be seen that both the meta-atoms produces strong Mie resonance in the collective modes. However, due to the strengthening and weakening effect of the in-phase and out-of-phase oscillation, it can be inferred that the intensities of electric fields are enhanced and diminished between two meta-atoms accordingly, especially in their gap region. The intensities of electric field in plane y=0 at two collective resonance frequencies are shown in figure 4(a) and figure 4(b). Both the meta-atoms have strong localized electric fields inside. However, due to the different coupling interactions of the two modes, their radiated fields are superimposed and counteracted, respectively. We have calculated the intensity of the electric fields on the central axis between two meta-atoms, as shown in figure 4(c) and 4(d). For the first red-shifted collective mode, the intensity of electric field has been enhanced by generally 4 to 5 times compared to the incident electric field in most regions on the central axis. While for the second blue-shifted collective mode, the intensity of electric field on central axis has been diminishes to 0 to 0.7 times compared to the incident one. This corroborates our hybridization model of dielectric metamolecule. Since most electric fields are localized inside two meta-atoms and the out-of-phase oscillation impels counteracting of the radiated fields at the gap between two meta-atoms, thus the electric fields are weakened compared to the incident fields. The certain zones between meta-atoms in two collective modes can be regarded as "hot zone" and "cold zone", respectively.

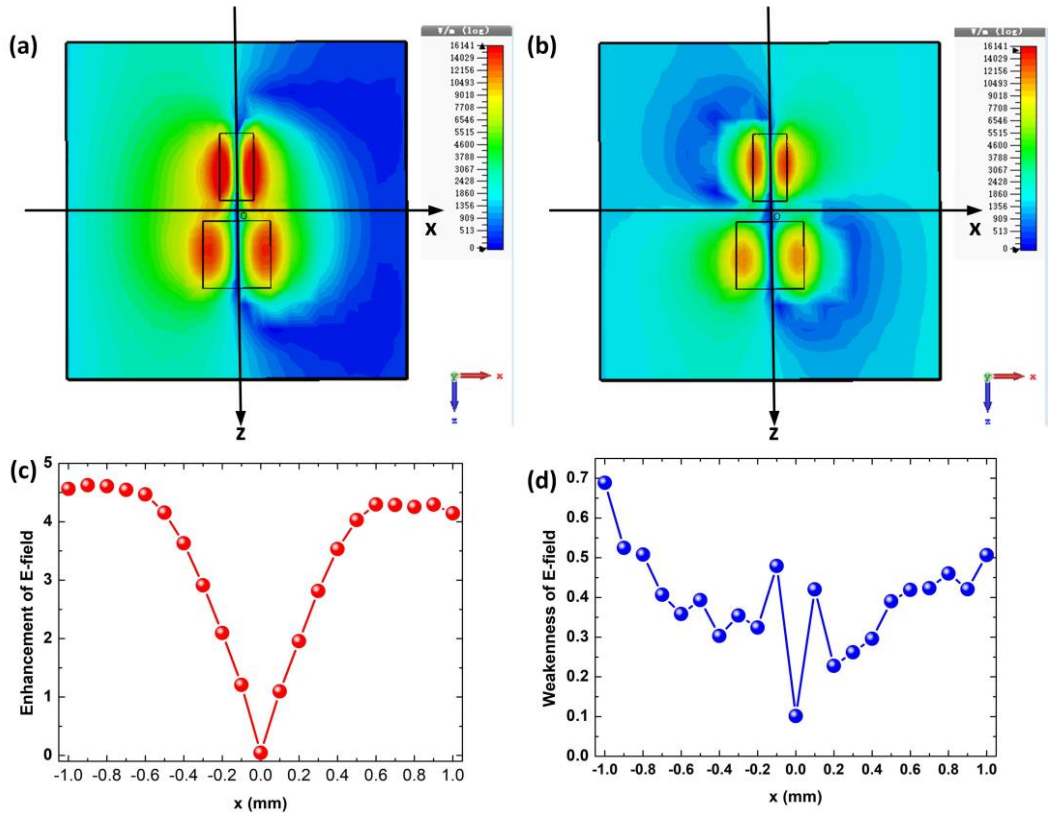

Figure 4. Electric field intensity of metamolecule at (a) 10.16GHz and (b) 10.727GHz. (c) Electric field enhancement $|E|/|E_0|$ at 10.16GHz and (d) electric field reduction $|E|/|E_0|$ at 10.727GHz on the central axis (y=z=0) with x changes from -1 to 1.

Utilizing the in-phase collective mode of the metamolecule, we can obtain "hot zone" with enhanced electric fields. Since the collective mode results from the coupling hybridization effect, factors that affect the coupling strength can be taken advantage to tune the enhancement amplitudes of the electric fields. As shown in figure 5(a), the intensity of E-field at fixed point (-0.8, 0, 0) is enhanced increasingly as the gap distance between two meta-atoms decreases. When the gap is 0.1mm, the first collective mode resonates at 9.79GHz, where the enhancement amplitude of electric field intensity reaches up to 7.4 times compared with the incident electric field. Besides, when the "hot zone" moves along the z-axis, the intensity can also been adjusted due to different coupling influence of the two "heteronuclear" meta-atoms. What is more, the enhancement amplitudes of E-field can be tuned in an even larger range by using meta-atoms with various permittivities, seen supplementary materials. Since the localized displacement electric fields inside meta-atoms in collective resonance modes have a great relationship with the permittivities of resonators.

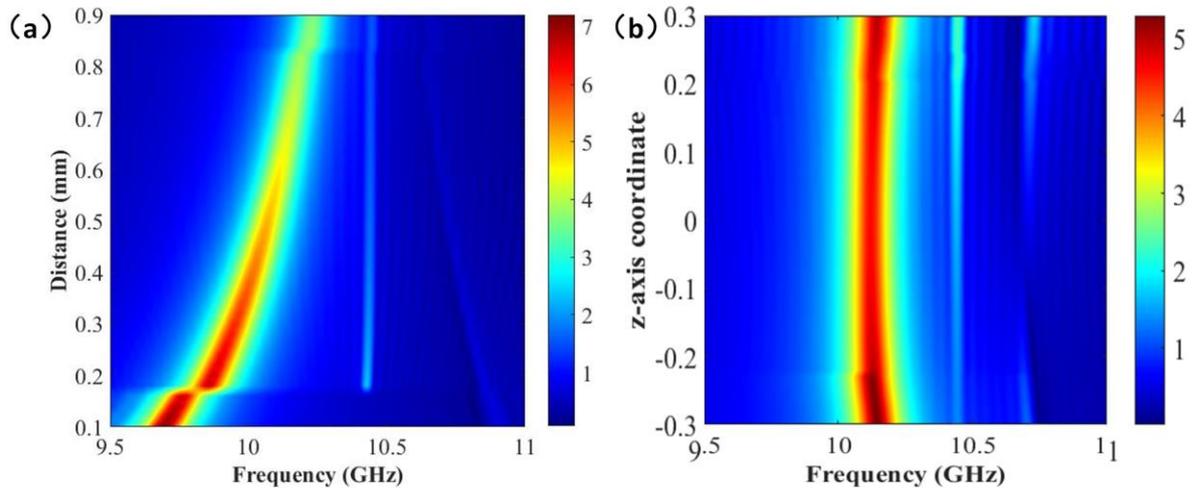

Figure 5(a) Electric field enhancement $|E|/|E_0|$ at point (-0.8,0,0) with gap increasing from 0.1mm to 0.9mm at 9.5-11GHz. (b) Electric field enhancement $|E|/|E_0|$ at (-0.8,0,z) at 9.5-11GHz, where position of z moves from surface of one meta-atom to another, that is, z changes from -0.3 to 0.3.

**4. Conclusion**

In this study, hybridization of the dielectric metamolecule induces transparent window as well as two new collective modes. At the frequencies of two collective modes, displacement electric currents distribute as magnetic loops with in-phase and out-of-phase oscillation in two meta-atoms, respectively. The superimposition and counteracting interactions result in strengthening and weakening effects of the electric fields. Besides, the enhancement amplitudes can be adjusted via coupling factors and resonator parameters. These special "hot zones" and "cold zones" in metamolecule in collective modes are promising for heating in microwave band and amplifying or shielding signals in communication fields.

**Acknowledgements**

This work was supported by the Basic Science Center Project of NSFC under grant No. 51788104, as well as National Natural Science Foundation of China under Grant Nos. 51532004 and 11704216.

**Conflicts of interest:** There are no conflicts to declare.